\RequirePackage{graphicx}
\documentclass[aps,pra,floatfix,reprint,longbibliography]{revtex4-2}
\usepackage{amssymb}
\usepackage{enumitem}
\usepackage{mathtools}
\usepackage{xcolor}

\usepackage{float}

\usepackage{graphicx}
\usepackage{natbib}
\usepackage{dcolumn}
\usepackage{bm}
\usepackage{mathtools}
\usepackage{braket}
\usepackage[utf8x]{inputenc}
\usepackage{amsmath}
\usepackage{multirow}
\usepackage{amsfonts}
\usepackage{changepage}
\usepackage{amssymb}
\usepackage{rotating} 
\date{Submitted: \today}

\usepackage{amsthm}

\theoremstyle{plain}
\theoremstyle{definition}

\begin{document}
\title{Relaxation of a Stationary State on a Quantum Computer Yields Unique Spectroscopic Fingerprint of the Computer's Noise}




\author{Scott E. Smart$^{1}$, Zixuan Hu${}^2$, Sabre Kais${}^2$, and David A. Mazziotti${}^{1}$ }
\email[]{damazz@uchicago.edu}
\affiliation{Department of Chemistry and The James Franck Institute, The University of Chicago, Chicago, IL 60637}%
\affiliation{Department of Chemistry, Department of Physics, and Birck Nanotechnology Center, Purdue University, West Lafayette, IN, 47907, United States}

%
\date{Submitted April 25, 2021}

\begin{abstract}
Quantum computing has the potential to revolutionize computing for certain classes of problems with exponential scaling, and yet this potential is accompanied by significant sensitivity to noise, requiring sophisticated error correction and mitigation strategies. Here we simulate the relaxations of stationary states at different frequencies on several quantum computers to obtain unique spectroscopic fingerprints of their noise.  Response functions generated from the data reveal a clear signature of non-Markovian dynamics, demonstrating that each of the quantum computers acts as a non-Markovian bath with a unique colored noise profile.  The study suggest that noisy intermediate-scale quantum computers (NISQ) provide a built-in noisy bath that can be analyzed from their simulation of closed quantum systems with the results potentially being harnessed for error mitigation or open-system simulation.
\end{abstract}



\maketitle

Quantum computing, as conceived by Feynman \cite{Feynman1982}, has the potential to revolutionize computing for certain classes of problems with exponential scaling in the physical and social sciences and engineering \cite{Abrams1999, Lloyd1996, Kandala2017, McArdle2020, Head-Marsden2020,  Farhi2014, Hu2020, Sager2020, Smart2021}. Central to the quantum computing paradigm is the quantum process of entanglement by which a pure-state quantum system develops a probability distribution over multiple classical outcomes. Entanglement allows us to process and store exponentially more information than a classical computer. This potential capability and its advantages, however, come with a significant sensitivity to noise \cite{Breuer2007, Clerk2008, Lidar2019, Head-Marsden2020} that introduces errors that degrade performance, especially on current-to-near-term quantum computers.  Significant advances have been made in the past decade in error correction and mitigation \cite{Krantz2019, Kandala2019, McArdle2019, Smart2019, Smart2021a, Endo2020}, but further advances are needed to not only understand noise but also to control noise for its mitigation or exploitation.  While a qubit can be characterized by its relaxation or dephasing times and a gate understood through quantum process tomography \cite{Mohseni2007, Mohseni2008, Nielsen2010}, such techniques are prohibitive for more than a few qubits.  Here we simulate the relaxation of a stationary state on a quantum computer to obtain the unique spectroscopic fingerprint of the quantum computer's noise.

We study the time evolution of a stationary state on a quantum computer.  Because the state is stationary, in the absence of noise the state will remain the same for all time.  Because the quantum computer is noisy, however, the quantum system begins to relax towards the ground state with the time evolution reflecting the noise of the computer.  The noise adds a bath to the time evolution of the state, transforming the stationary closed system into a time independent open quantum system~\cite{Breuer2007, Lidar2019}.  Moreover, by analyzing the time dependence of the relaxation, we obtain unique spectroscopic signatures of the noise on the IBMQ Experience quantum computers. Analysis of the results reveals significant non-Markovian behavior~\cite{Breuer2007, Lidar2019, Head-Marsden2019a} on the computers. Furthermore, the results show that stationary-state quantum simulation can be effectively applied to characterize noise on quantum computers. Such studies promise to provide new directions for not only mitigating but also utilizing existing noise for open-system quantum simulations~\cite{Tseng2000, Bacon2001, Wang2011, Sweke2015, Hu2020, Endo2020a, Hu2021, Head-Marsden2021, Kamakari2021}.

Time evolution of a stationary state prepared on a noisy quantum computer can be described by the equation of motion of the density matrix $D$~\cite{Breuer2007, Lidar2019, Head-Marsden2019}
\begin{equation}
 \frac{dD}{dt} = -\frac{i}{\hbar} [{\hat H}, D] +\int_0^t \mathcal{K}(t,\tau)D(t,\tau) d \tau,
 \label{norm}
\end{equation}
where $\hat H$ is the Hamiltonian operator of the stationary state and $\mathcal{K}(t,\tau)$ is the memory kernel representing the quantum computer's noise.  In the limit that the noise on the quantum computer vanishes, the memory kernel vanishes and the equation simplifies to the quantum Liouville (von Neumann) equation.  If the initial state is a stationary state of ${\hat H}$, all of the non-trivial time dependence results from the noise. Normal noise spectroscopy or characterization of a system could be performed by using a number of techniques, such as with Rabi spectroscopy, or with a tunable system \cite{Yoshihara2014, Norris2016, Yan2013, Schoelkopf2003, Ithier2005, Bylander2011}.  Here, to probe the frequency dependence of the noise, we consider the family of scaled Hamiltonians ${\hat H} = \hbar \omega {\hat O}$ where ${\hat O}$ is a dimensionless operator.  By changing $\omega$, we can control the energy difference between the ground and first excited state of the Hamiltonian.  The time-dependent response of the system to different values of $\omega$ provides us with spectral information about the noise on the quantum computer.  Importantly, we are attempting to simulate the solution of the quantum Liouville equation on the quantum computer---dynamics with the memory kernel set to zero.  Consequently, all deviations from the closed-system evolution are originating from the noise of the given quantum computer which creates without our direction an effective memory kernel for the time evolution.

We begin with a single qubit system with the Hamiltonian matrix $H_0(\omega) = \omega \sigma_z$ where we use atomic units with $\hbar = 1$. We prepare the system in the excited state $|\psi\rangle = |1\rangle$ and evolve the system according to $\exp [{-i H \tau }] = R_z(-\omega \tau)$, using repeated single qubit gates with the results shown in Fig.~1. The time step is arbitrary to the extent that it can be rescaled with the strength of $V$ and $\omega$, and hence, we set $\tau=\frac{1}{3}$, which serves to highlight system-bath interactions when the frequency $\omega \in [0,1]$.  If we allowed the system to relax without applying any gates, this would essentially be a $T_1$ experiment (where we could use the physical gate times), measuring the relaxation time for an excited state. However, the noise sources here, represented by the non-vanishing memory kernel, generate non-Markovian behavior. In Fig.~1 the non-Markovian behavior can be seen from the oscillations in the population of the ground state, which reveal a memory dependence beyond the pure decay of Markovian dynamics.  Furthermore, the oscillations are more pronounced at lower frequencies, indicating a bath with colored noise.


\begin{figure}[h]
\includegraphics[scale=0.22]{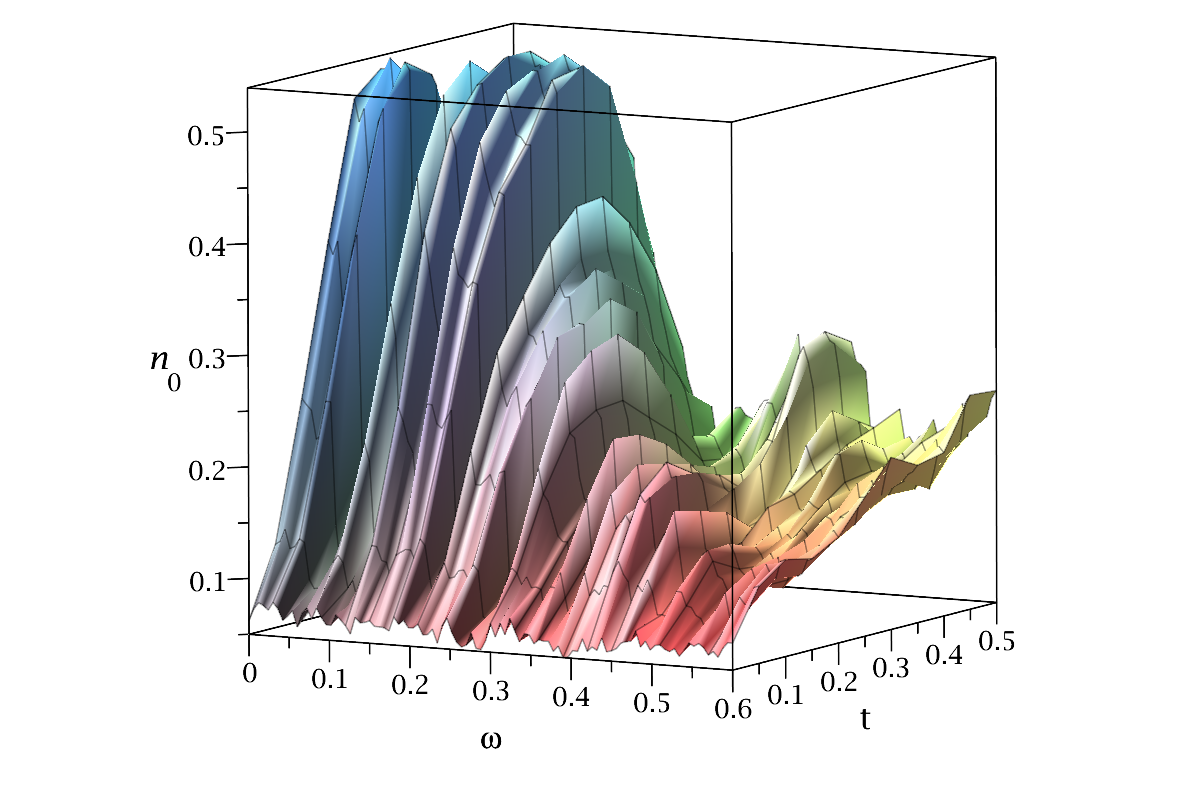}
\caption{Scan of population of the $ | 0 \rangle$ state $n_0$ as a function of time and frequency $\omega$ on the IBMQ Armonk device. We initially prepare the state in the $1$ state and then apply 100 total gate sequences of $\exp[{-i\tau H_0}]$, where $\tau=\frac{1}{3}$, $H_0=\omega  \sigma_z$, and $\omega \in (0,0.6)$. The time $t$ is equal to the number of gates times $\tau$ scaled by 0.015. Effectively, non-Markovian behavior is manifest in the oscillatory trends present across all frequency ranges, which slowly tapers down with the application of higher frequencies. The appropriate resolution is difficult to determine, and the wells are not stochastic, as a purely stochastic phenomena would not be continuous along the time series, as each time series is sampled independently.}
\end{figure}

The spectral density with respect to a quantum noise source $A$ can be characterized as:
\begin{align}
    S_{A}(\omega) &= \int_{-\infty}^{\infty} d\tau \exp[{i\omega \tau}] \sum_{\alpha\beta} \rho_{\alpha \alpha}\langle \alpha | A(\tau)| \beta  \rangle \langle \beta | A(0)| \alpha \rangle \nonumber \\
    &= 2\pi \sum_{\alpha,\beta} \rho_{\alpha \alpha} |V_{\alpha\beta}|^2 \delta( \epsilon_\beta-\epsilon_\alpha - \omega ),
\end{align}
where $\alpha$ and $\beta$ are energy eigenstates of $H_0$ and $\rho$ represents elements of the density matrix. The spectral density can be related to the rate of population change through first-order perturbation theory with the well-known Fermi's golden rule \cite{Clerk2008, Schoelkopf2003, Tokmakoff2014}. Figure~2 is constructed from Fig.~1, showing the rate of change in the initial part of the time evolution as a function of frequency, which provides a rough spectrum of the noise on the quantum computer from the perspective of the simulated system.  Note that for frequencies above 0.6 (relative to a value of $\tau=1/3$), the noise profile shows low intensity signals, likely from thermal noise.

\begin{figure}[h]
\includegraphics[scale=0.45]{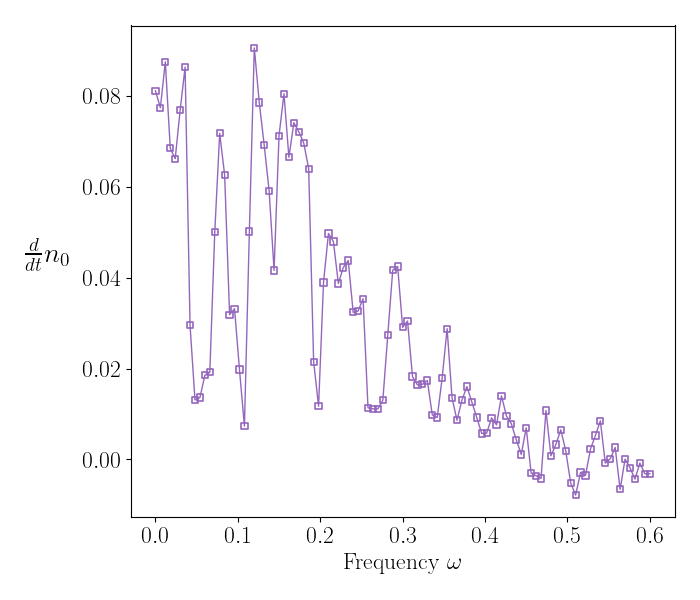}
\caption{Rate of ground state populations changes across the frequency spectrum for the single qubit system taken after the first time step. }
\end{figure}

The significant portion of the noise can be ascribed to a transverse noise source ($\sigma_x, \sigma_y$) which allows for transitions between the ground and excited states. The quantum devices we studied utilize a sequence of elementary gates denoted as $U_1$, $U_2$, and $U_3$ gates. In particular, these gates consist of alternating frame changes ($R_z$ gates) and rotations (X90 rotation, $R_x$ gates), where $U_k$ contains $k$ frame changes and $k-1$ transverse rotations. If only the $U_1$ gate is used to model $\exp[-i\omega\sigma_z\tau]$, then a pulse is not applied, and the gate applications correspond to a standard $T_1$ experiment, measuring the relaxation of the excited state. However, here we specifically use $U_3$ gates, which are generic single qubit rotations, and which alternate between transverse and longitudinal rotations. In fact, one way to model the observed behavior in the qubit system is to apply a constant $\exp [i\theta \sigma_x]$, where $\theta$ is taken to be a small angle, following every gate application. Yet, this does not account for the troughs seen within the range from 0.03 to 0.3, which are not stochastic effects. Whether or not the noise channels are a result of an improper calibration of the $X90$ gate resulting in a systematic over rotation, or are part of a  completely different noise source, becomes irrelevant for the noisy system in that the quantum system could be equivalently described from either perspective \cite{Yan2013,McKay2018}. For a user or algorithm that does not control the qubit and gate level of the simulation, we posit that the effects of the two are the same, and the quantum system could be viewed as experiencing the same.

Next we also investigated the trajectories beyond the populations, such as how the quantum state vector moves in the Bloch sphere. For frequencies in the high region of the spectral density and not allowing for coupling between the bath and system, a slow precession around the axis corresponding to $H_0$ is observed. However, for frequencies which couple to the bath, the system can be strongly pulled around the Bloch sphere resulting in interesting trajectories. We model some of the trajectories in Fig.~3. We also investigated these instances with systems described by $H_0 = \omega \sigma_x$ and $H_0 = \omega \sigma_y$, which are included in the Supplemental Information, and found essentially the same general behavior. We also found that these trends were (as expected) not the same across each device, although a similar noise signal could be identified. We identified similar resonant regions with other devices, albeit with the strength of the various coupling regions depending on the device.

\begin{figure}[h]
\begin{center}
\includegraphics[scale=0.30]{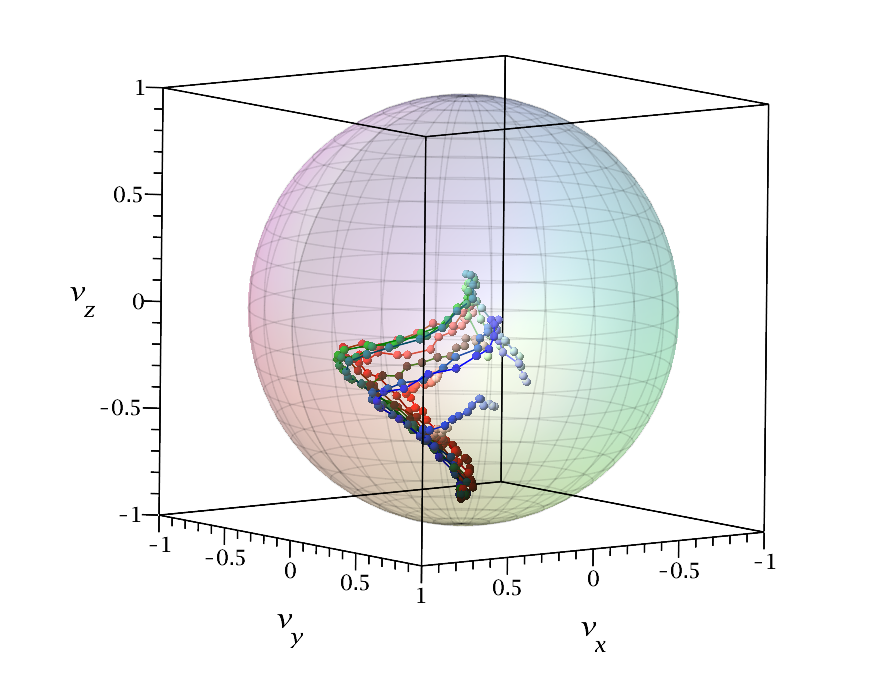}
\end{center}
\caption{Scan of several trajectories represented in the Bloch sphere for a single qubit device within a frequency of $\omega = 0.05$ and $0.15$ in increments of $0.01$. $v_i$ represents the value of $\langle \sigma_i\rangle $ for a given state. Each simulation had an equal number of gate applications and changed only the parameter $\omega$ in $\exp [-i\tau H ]= \exp [-i\frac{\omega \sigma_z}{3}]$. $\omega $ here decreases from red ($\omega=0.05$) to green ($\omega=0.10$) to blue ($\omega=0.15$).}
\end{figure}

By identifying frequencies of interest on the qubits themselves, we can construct systems which respond uniquely to the bath, allowing for selective transitions between different eigenstates of the system. With this in mind we simulated a two-qubit system with a Hamiltonian defined by:
\begin{equation}
    H(\omega_1,\omega_2) = \omega_1 \sigma_z^1 + \omega_2 \sigma_z^2 .
\end{equation}
This Hamiltonian has the computational basis as its energy eigenbasis, and we can describe the energies as:
\begin{align}
    E_{qr} = (-1)^{q}\omega_1 + (-1)^{r}\omega_2,
\end{align}
where $q$ and $r$ are elements of the computational basis $0$ and $1$. Transitions between two states can then be written as:
 \begin{equation}
     E_{qr} - E_{st} = 2\omega_1 (1-\delta^q_s) (-1)^q + 2\omega_2 (1-\delta^r_t) (-1)^r.
 \end{equation}
Changing a frequency $\omega_i$ then primarily influences transitions on qubit $i$. By scanning over single qubit frequencies, we can obtain a simple noise profile, and then choose appropriate $\omega_i$ to influence the system. Using these frequencies we highlight four different cases in Fig.~4, showing interacting and non-interacting frequencies for each qubit. Additionally, we presented the calculated transition rates for each population in Table I.

\begin{figure}[h]
\begin{center}
\includegraphics[scale=0.5]{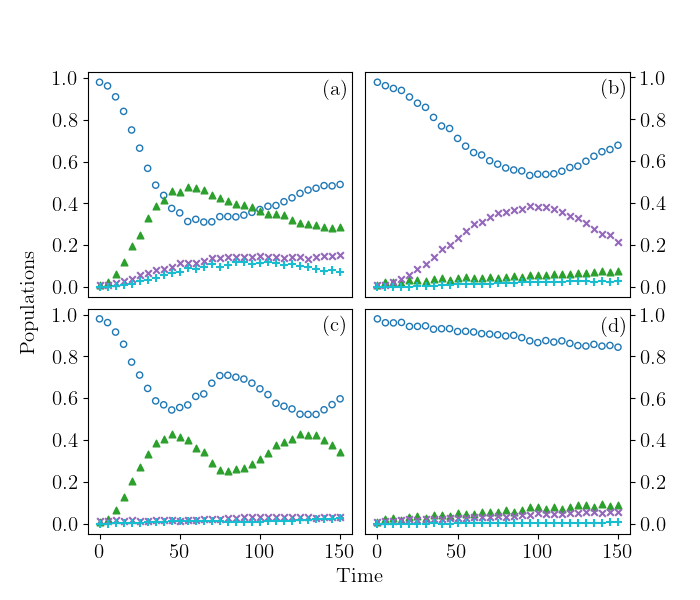}
\end{center}
\caption{Populations of a two qubit system undergoing time evolution implemented as $\exp -i\tau H_0$, where $\tau=1/3$. The system Hamiltonian can be described with two frequencies: $H(\omega_1,\omega_2)= \omega_1 \sigma_z^1 + \omega_2 \sigma_z^2 $. The different plots describe permutations of two different values of $\omega_1$ and $\omega_2$: (a), $(\frac{1}{20},\frac{1}{20})$; (b): $(\frac{1}{20},\frac{1}{2})$; (c): $(\frac{1}{2},\frac{1}{20})$ and; (d): $(\frac{1}{2},\frac{1}{2})$. Note that the bath is not symmetric with respect to swapping the qubits, as there is a stronger interaction with the second qubit as compared to the first qubit. The populations are represented as $n_{00}:\circ$ (blue), $n_{01}:\triangle$ (green), $n_{10}:\times$ (purple), and $n_{11}:+$ (cyan).}
\end{figure}

\begin{table}[]
    \caption{Rate of population change shortly after initialization for the two qubit system depicted in Fig.~4, where $H=\sigma_z^1 \omega_1 + \sigma_z^2 \omega_2 $, and $\tau=\frac{1}{3}$. Measurements were taken every 15 gate applications. The strongest bath interaction is related to low frequency $\omega_1$, which when isolated or in conjunction with $\omega_2$ has strong enough coupling to induce state transitions. The only significant increase in the $n_{11}$ state occurs when a low transition frequency is available for both qubits. Additionally, these rates are substantially lower than what is observed in the single qubit system, which corresponds with the better gate fidelities for this system (see Supplementary Information).}
    \centering
    \begin{tabular}{cc||cccc}
& & \multicolumn{4}{c}{$\frac{d}{dt}n_i ~(\times 10^{-3})$} \\
$\omega_1 $    & $\omega_2 $   & $n_{00} $&  $n_{01} $ & $n_{10} $ & $n_{11}$ \\ \hline
$\frac{1}{20}$   & $\frac{1}{20}$   &  $-16.4$ & $12.4$ & $2.5$ & $1.5$ \\
$\frac{1}{20}$    &  $\frac{1}{2}$ & $- 4.7$ & $0.5$ & $4.1$ & $0.2$ \\
 $\frac{1}{2}$  &  $\frac{1}{20}$ & $-13.7$ & $13.5$ & $0.0$ & $0.2$ \\
$\frac{1}{2}$ &  $\frac{1}{2}$ & $-1.1$ & $0.7$ & $0.4$ & $0.0$ \\
\hline
    \end{tabular}
    \label{tab1}
\end{table}

The system here demonstrates asymmetry between the two qubits, with stronger coupling present on the second qubit. For large $\omega_1$ and $\omega_2$, which do not strongly couple the bath and qubit system, we witness a region of linear decay towards a uniformly depolarized state. For smaller $\omega_1$ and $\omega_2$, we clearly have a very dynamic system, allowing for transitions amongst all four states. The same principles can be applied to other systems, i.e. with more complex two-qubit Hamiltonians involving CNOT gates.

By characterizing noise properties of the system, we may be able to design better error mitigation techniques or novel approaches to the simulation of open quantum systems where the quantum computer's noise is harnessed as an effective bath.  For example, examining the spectral profile of the bath from the simulation of stationary quantum states may provide a unique spectroscopic fingerprint of the quantum computer.  With such a fingerprint we may be able to design simulation algorithms that account for this fingerprint, providing a potentially elegant approach to error mitigation for real-world applications.  Furthermore, there may be important scenarios such as open-quantum-system simulation where the presence of noise may be a beneficial quantum resource.  To model an open quantum system, we may be able to use the quantum computer's noise to represent a significant part of the model's bath.  Use of the quantum computer's inherent noise could potentially permit the simulation of an open quantum system at a significantly reduced computational cost in terms of both gate and qubit resources.

Additionally, although noise sources in driven evolution cannot typically be attributed to singular sources, the absence of such specificity is not necessarily an issue for practical quantum computing.  As mentioned above, characterizing whether the transverse signal is a systematic over-rotation or an errant noise source is useful in calibration but not so important for end-user applications. For a complex quantum simulation, the effects of a single source of error (unless uniquely distinct) cannot be easily distinguished amidst the entire chorus of potential noise sources. From the complex set of instructions on a quantum computer emerges a complex noise profile, which is manifest in the difficulty of simulating multi-qubit noise phenomena.  Through a device specific approach, we can utilize the effective bath's spectroscopic information to design more device-specific techniques and algorithms that could improve future applications.

In this work we simulate the time evolution of a stationary quantum state on a noisy quantum computer.  Noise, we show, causes the stationary state to relax and evolve with time.  Furthermore, spectroscopic analysis of this time evolution provide a frequency spectrum---a spectroscopic fingerprint---of the noise on the quantum computer.  Understanding the noise profile may allow us to create parameterized systems in which we influence state transitions with the quantum device serving as a non-Markovian bath.  These ideas provide a first step towards harnessing the unique quantum noise of each quantum computer in novel approaches to error mitigation and noise-related simulations such as the simulation of open quantum systems.

\noindent\textbf{Author contributions.} D. M. and S. K. conceived of the project.  S. S., Z. H., S. K. and D. M. developed the theoretical framework and designed the computations.  S. S. wrote the code and performed the computations.  S. S., Z. H., S. K. and D. M. analyzed the results and wrote the paper.

\noindent\textbf{Data availability.} Data will be made available upon reasonable request.

\noindent\textbf{Code availability.} Code will be made available on a public Github repository upon publication.

\section*{Acknowledgements}
We acknowledge the financial support from the U.S. Department of Energy (Office of Basic Energy Sciences) under Award No. DE737 SC0019215. D. A. M. and S. K. also acknowledge the support of the National Science Foundation under award numbers CHE-1565638, CHE-2035876, DMR-2037783, and CHE-1955907. The views expressed are of the authors and do not reflect the official policy or position of IBM or the IBM Q team.

\section*{Methods}
In each simulation we used atomic units, and the time steps were relative to meaningful scales on the quantum device. While we could associate the results to a physical time through the known gate lengths, we focus on presenting the time evolution from the perspective of the simulated system, which could have arbitrary energies and time values, and which ultimately is beholden to the gate errors. As mentioned in the text, the system Hamiltonian have single qubit or in the two-qubit system, a sum of two-single qubit gates, both of which can be implemented as exact exponentials. These are implemented with $U_3$ gates on the quantum computer, which have the form:
\begin{align}
U_3(\theta,\phi,\lambda) &= R_z(\phi) R_x (-\frac{\pi}{2})R_z(\theta) R_x (\frac{\pi}{2}) R_z(\lambda) \\ &=
\begin{pmatrix}
\cos (\frac{\theta}{2})  & -e^{i\lambda}\sin (\frac{\theta}{2}) \\
e^{i\phi} \sin (\frac{\theta}{2}) & e^{i(\phi+\lambda)}\cos  (\frac{\theta}{2})
\end{pmatrix}.
\end{align}
Each circuit was prepared and then measured $2^{13}$ times at each evaluated point. The simulations used cloud-available quantum devices accessible through IBM Quantum Experience. The particular results reported here were performed on Armonk (single qubit device) and Rome (5 qubit device) over several days. The devices use fixed-frequency transmon qubits with co-planer waveguide resonators~\cite{Koch2007,Chow2011}. We use the Python package Qiskit (v 0.15.0) \cite{Qiskit} to interface with the device. Specific device properties relevant to each run can be found in the Supplemental Information.

\bibliography{
    biblio/b0_oqs.bib,
    biblio/b1_quant_sim.bib,
    biblio/b2_quant_misc.bib,
    biblio/b3_supp.bib,
    biblio/b4_books.bib,
    biblio/QCFingerprint.bib
}

\end{document}